\documentclass[twocolumn,showpacs,pra,amsmath,amssymb,floatfix,eqsecnum]{revtex4-1}
\renewcommand{\onlinecite}{\cite}
\usepackage{graphicx}   
\usepackage{bm}

\usepackage{enumerate}

\newcommand{\pd}[2]{\frac{\partial #1}{\partial #2}}

\newcommand{\gc}{\mathcal{W}}

\DeclareMathOperator{\Tr}{Tr}
\newcommand{\ut}{\tilde{u}}

\newcommand{\pla}{\partial_{\Lambda}}

\newcommand{\gal}{\Gamma_{\Lambda}}

\newcommand{\gwl}{\Gamma_{\Lambda}}

\newcommand{\dd}{\mathrm{d}} 
\newcommand{\ee}{\mathrm{e}}

\begin{document}

\title{The zero-dimensional $O(N)$ vector model as a benchmark for perturbation theory, the large-$N$ expansion and the functional renormalisation group}

\author{Jan Keitel and Lorenz Bartosch}

\affiliation{Institut f\"{u}r Theoretische Physik, Universit\"{a}t
  Frankfurt,  Max-von-Laue Stra{\ss}e 1, 60438 Frankfurt, Germany}

\date{November 14, 2011}

 \begin{abstract}

We consider the zero-dimensional $O(N)$ vector model as a simple example to calculate $n$-point correlation functions using perturbation theory, the large-$N$ expansion, and the functional renormalisation group (FRG). 
Comparing our findings with exact results, we show that perturbation theory breaks down for moderate interactions for all $N$, as one should expect. 
While the interaction-induced shift of the free energy and the self-energy are well described by the large-$N$ expansion even for small $N$, this is not the case for higher-order correlation functions.
However, using the FRG in its one-particle irreducible formalism, we see that very few running couplings suffice to get accurate results for arbitrary $N$ in the strong coupling regime, outperforming the large-$N$ expansion for small $N$. We further remark on how the derivative expansion, a well-known approximation strategy for the FRG, reduces to an exact method for the zero-dimensional $O(N)$ vector model.

\end{abstract}

\pacs{
11.10.Hi, 
11.10.Jj, 
11.15.Pg 
}

\maketitle
\section{Introduction}

Field theory and its applications to condensed matter and high-energy physics are difficult subjects in the natural sciences. The solution of almost any field theory requires approximations which rely on the mastery of complicated expressions such that it is easy to get bogged down in technical details. As there is no exact solution to compare with, the quality of results from different approximation strategies like perturbative methods, the large-$N$ expansion, or non-perturbative methods such as the functional renormalisation group (FRG) is often difficult to judge. 

An instructive problem to learn about the conceptual underpinnings of asymptotic series (see e.g. Refs.~\onlinecite{Schulman81,Negele88,Altland10}) and Feynman diagrams (see e.g. Ref.~\onlinecite{Zee10}) in a simple context is the evaluation of the integral (with $r$ real and $u>0$)
\begin{equation}
  \label{eq:ZofJ}
  \mathcal{Z}(J) = \int_{-\infty}^{\infty} \dd\varphi\, \exp\left( 
    - \frac{r}{2} \varphi^2 - \frac{u}{4!} \varphi^4
+ J \varphi \right) \;.
\end{equation}
For $u = 0$ and $r>0$, this integral reduces to a Gaussian integral which can be calculated exactly, resulting in $\mathcal{Z}_0(J) = \sqrt{{2 \pi}/{r}} \exp\left(J^2/2r \right)$.
Keeping $u$ finite, $\mathcal{Z}(J)$ corresponds to the classical partition function of a particle in an anharmonic potential.
Alternatively, we can also think about this partition function as a toy model of a (zero-dimensional) field theory.
By taking derivatives of $\mathcal{Z}(J)$ with respect to $J$, it is possible to obtain all $n$-point functions (also called correlation functions or Green functions),
\begin{equation}
  \label{eq:GreenFunctions}
  G^{(n)} 
  = \frac{1}{\mathcal{Z}(0)} \frac{\partial^n \mathcal{Z}(J)}{\partial J^n}  \Biggr\rvert_{J=0}\;.
\end{equation}
Since the action $\mathcal{S}(\varphi) = \frac{r}{2} \varphi^2 + \frac{u}{4!} \varphi^4$ is invariant under the $Z_2$ transformation $\varphi \to -\varphi$, all  $n$-point functions with odd $n$ vanish.
In fact, for the zero-dimensional field theory considered here, all $n$-point functions can also be calculated by taking successive derivatives of $\mathcal{Z} = \mathcal{Z}(0)$ with respect to $r$, but as this feature does not generalise to a generic field theory, we follow here the usual strategy of including the source term $J \varphi$ and taking derivatives with respect to the sources.
It turns out that the integrals determining the $n$-point functions $G^{(n)}$ can all be evaluated analytically (see e.g. Ref.~\onlinecite{Schulman81} for the case $n=0$).
However, in order to learn about perturbative many-body techniques, it is instructive
to expand Eq.~(\ref{eq:ZofJ}) or Eq.~(\ref{eq:GreenFunctions}) for $r > 0$ in powers of $u$. 
As one should expect, such a perturbative approach is justified for small $u/r^2$ and breaks down for $u/r^2 = \mathcal{O}(1)$. 

In order to go beyond simple perturbation theory one usually has to sum up an infinite series of diagrams. But which classes of diagrams should be included? What if there is no small parameter to expand in? A useful strategy is to duplicate an internal field (in our case $\varphi$) to an $N$-component field and to calculate quantities in powers of $1/N$. This approach is called the large-$N$ expansion. By including the first one or two terms, one hopes to obtain reliable results even for small $N$.
Generalising the integral in Eq.~(\ref{eq:ZofJ}) to an $N$-dimensional integral leads to a zero-dimensional $N$-component field theory, which is also known as the zero-dimensional $O(N)$ vector model \cite{Schelstraete94}. 
As we will show below, the extrapolation of the leading-order large-$N$ result to the cases $N = 1$ or $N=2$ gives rise to sizeable deviations from the exact result. 
While it is possible to derive higher-order coefficients in a series expansion in $1/N$ for the zero-dimensional $O(N)$ vector model, this is usually problematic in more involved field theories. Clearly, there is a need for alternative methods which allow for a non-perturbative solution of field theories.

A non-perturbative method to determine all $n$-point functions 
is the functional renormalisation group (FRG).
There are basically two different implementations of the FRG (also called non-perturbative RG):
The Wilson-Polchinski approach \cite{Polchinski84} and the effective average action approach \cite{Wetterich93,Bonini93,Morris94}. 
While the Wilson-Polchinski approach was used with much success in proofs of renormalisability \cite{Polchinski84,Keller91}, the effective average action approach is better suited for most practical purposes  \cite{Morris98,Bagnuls01,Berges02,Gies06,Pawlowski07,Delamotte07,Kopietz10,Metzner11,Braun11}. The reason for the success of the latter is grounded on the fact that it is based on the generating functional of one-particle irreducible vertices (also known as the effective action).

The evaluation of the partition function $\mathcal{Z}(J)$ given in Eq.~(\ref{eq:ZofJ}) has already served as a simple example for an application of the FRG \cite{Schoenhammer01,Meden03,Kopietz10,Pawlowski10} (see also Ref.~\cite{Salmhofer07}).
Within the FRG, a regulator is added to the Gaussian part of the action. This regulator is usually implemented in such a way that it gives an extra mass to low-energy fluctuations. For the zero-dimensional $O(N)$ vector model it effectively replaces the mass term $r$ by $r+R_\Lambda$, where initially $R_\Lambda \to \infty$ such that the dimensionless parameter $u/(r+R_\Lambda)^2 \to 0$ and the initial condition can be determined exactly. 

The central quantity to compute within the framework of the one-particle irreducible FRG is the effective action. This effective action is the generating functional of all one-particle irreducible vertices (for the zero-dimensional $O(N)$ vector model all functionals reduce to functions).
All $n$-point functions can be derived from this generating functional.
In the presence of a regulator, the effective action is replaced by an effective average action which for $R_\Lambda \to \infty$ turns out to be equal to the action itself.
Gradually removing the regulator, the flow of the effective average action is governed by a flow equation \cite{Wetterich93, Bonini93,Morris94} which is simple, but nevertheless exact. During the flow, an infinite number of couplings is generated, and trying to determine the exact solution is usually a hopeless endeavour \cite{Schuetz05}.
Finding a good truncation by keeping some couplings, but neglecting others, is maybe the most difficult step in applying the FRG. 
However, as we will see, for the zero-dimensional $O(N)$ vector model, already very few coupling constants can be sufficient to obtain a very good description of the strong-coupling limit. 

The remainder of this paper is organised as follows: In Sec.~\ref{sec:ONVector} we generalise Eq.~(\ref{eq:ZofJ}) to the zero-dimensional $O(N)$ vector model, introduce its generating functions and discuss its symmetries. After presenting an exact solution for all $n$-point functions in Sec.~\ref{sec:ExactSolution}, we demonstrate how to calculate the $n$-point functions perturbatively in Sec.~\ref{sec:PerturbationTheory}. 
Some knowledge about the linked-cluster theorem and Wick's theorem are helpful here. We then discuss different versions of the large-$N$ expansion in Sec.~\ref{sec:largeN}. 
Higher-order results for both perturbation theory and the large-$N$ expansion are deferred to the appendix.
A brief introduction to the FRG is given in Sec.~\ref{sec:FRG} \cite{Delamotte04b},
where we also study the zero-dimensional $O(N)$ vector model by applying two widely used approximation strategies: While only three coupling constants suffice to give quite reliable results for the free energy and the two-point function within the vertex expansion, for the simple problem considered  here, the FRG flow equations can in principle be calculated by directly solving a partial differential equation numerically. In fact, another commonly used approximation strategy, the derivative expansion (which for the zero-dimensional $O(N)$ vector model has no derivative term), reduces to an exact method. 
Finally, we discuss and compare our results in the conclusion in Sec.~\ref{sec:Conclusion} and make a few further remarks on how the studies of this paper generalise to more involved field theories.

\section{Generating functions and symmetries of the zero-dimensional $O(N)$ vector model}
\label{sec:ONVector}

Generalising Eq.~(\ref{eq:ZofJ}) to an $N$-dimensional integral and  defining $\int \mathcal{D}\vec{\varphi} := \prod_{i=1}^N \int_{-\infty}^{\infty} \dd \varphi_i$, the generating function of all Green functions reads
  \begin{equation} \label{eq:ZofJ_N}
    \mathcal{Z}(\vec{J}) = \int \mathcal{D}\vec{\varphi} \, \exp \left( - \mathcal{S}(\vec{\varphi} ) + \vec{J}  \cdot \vec{\varphi} \right)\;.
  \end{equation}
Here, 
\begin{equation}
  \label{eq:action}
  \mathcal{S}(\vec{\varphi}) = \mathcal{S}_0 (\vec{\varphi}) + \mathcal{S}_{\text{int}} (\vec{\varphi}) = \frac{r}{2} \vec{\varphi}^2 + \frac{u}{4!} (\vec{\varphi}^2)^2 
\end{equation}
is a generalisation of the action given in Eq.~(\ref{eq:ZofJ}) and $\vec{\varphi}^2=\sum_{i=1}^N \varphi_i^2$ is the squared norm of the vector $\vec{\varphi}$. 
While only $\textrm{Re}\,(u)$ has to be positive for the integrals to converge, perturbation theory also requires $\textrm{Re}\,(r) > 0$. 
Even though both $r$ and $u$ can be considered to be complex, we will be mainly interested in the case where both of these parameters are real.
However, let us mention here that the exact and perturbative results to be derived in the following two sections are also valid for complex $r$ and $u$ (provided $\textrm{Re}\,(u) > 0$ for reasons of convergence and also $\textrm{Re}\,(r) > 0$ in the perturbative case).
For real $r \neq 0$, it is in principle possible to rescale $\vec{\varphi} \to \vec{\varphi}/\sqrt{|r|}$, thereby replacing $r$ by $\pm 1$ and $u$ by the dimensionless parameter $u/r^2$. However, let us not do so. We thereby avoid having to distinguish the cases $r<0$, $r=0$, and $r > 0$ (in which we will be mainly interested). Furthermore, $1/r$ can be interpreted as the free propagator, and we would rather not replace this free propagator by one.

The generating function $\mathcal{Z}(\vec{J})$ stores the information about all $n$-point functions, albeit for many purposes not in its most convenient form.
The generating function of all {\em connected} Green functions is obtained by simply taking the logarithm of Eq.~(\ref{eq:ZofJ_N}) (for a neat proof based on the replica trick, see e.g. Ref.~\onlinecite{Negele88}),
 \begin{equation} \label{eq:gf_gc}
     \mathcal{W} (\vec{J}) = \ln \left( \frac{\mathcal{Z}(\vec{J})}{\mathcal{Z}_0}\right) \;.
  \end{equation}
The normalisation with the partition function $\mathcal{Z}_0 = \left. \mathcal{Z}(\vec{J}=0)\right|_{u=0} = (2 \pi/r)^{N/2}$ of the free system lends $\mathcal{W}(0)$ the interpretation of the negative interaction-induced shift of the free energy.
Calling the expectation value of the field $\vec{\varphi}$
  in the presence of sources
  \begin{equation}
    \vec{\phi} = \langle \vec{\varphi} \rangle_{\vec{J}} = \pd{\mathcal{W}}{\vec{J}}\;,
  \end{equation}
we can perform a Legendre transformation to make the transition from the generating function of connected Green functions to the generating function of one-particle irreducible vertices \cite{Negele88},
  \begin{equation}
    \Gamma (\vec{\phi}) = \vec{J}(\vec{\phi}) \cdot \vec{\phi} - \gc \left(\vec{J}(\vec{\phi})\right) \;.
  \end{equation}
This is similar to using the Legendre transformation to switch from the Helmholtz free energy $F$ to the Gibbs potential $G$, which often turns out to be the more useful quantity.

    As a consequence of the generalisation of the simple integral in Eq.~(\ref{eq:ZofJ}) to $N$ fields, an $n$-point Green function now carries $n$ indices $i_1, \ldots, i_n$, ranging from 1 to $N$. 
Denoting the average with respect to the full action by
    \begin{equation} \label{eq:full_average}
    	\langle \ldots \rangle := \frac{\int \mathcal{D}\vec{\varphi}\,\ee^{-S(\vec{\varphi})} \; [\ldots]}{\mathcal{Z}}\;,
    \end{equation}
the $n$-point functions can be defined as
\begin{equation}
  \label{eq:GreenFunctionsN}
  G_{i_1,\dots,i_n}^{(n)} = \langle \varphi_{i_1} \dots \varphi_{i_n} \rangle 
  = \frac{1}{\mathcal{Z}} \frac{\partial^n \mathcal{Z}(J)}{\partial J_{i_1} \dots \partial J_{i_n} }  \Biggr\rvert_{J=0}\;.
\end{equation}
It turns out that for the present model, any (non-vanishing) Green function of order $n$ already determines all other Green functions of the same order.
To prove this statement, let us note that, by construction, our action
given in Eq.~\eqref{eq:action} is invariant under $O(N)$ rotations and therefore (as there are no derivative terms in zero dimensions)
depends only on $\vec{\varphi}^2$.
Since the action does not favour any particular direction in the $N$-dimensional space, the same also has to be true for all generating functionals. We therefore have $\mathcal{Z}(\vec{J}) = \tilde{\mathcal{Z}}(\vec{J}^2)$,
$\mathcal{W}(\vec{J}) = \tilde{\mathcal{W}}(\vec{J}^2)$, and $\Gamma (\vec{\phi}) = \tilde{\Gamma} (\vec{\phi}^2)$.
A more formal proof of the fact that any symmetry of the action is also
    a symmetry of the generating functionals can be found, for example, in Refs.~\onlinecite{Gies06,Kopietz10}.
It follows that
\begin{align}
    & \mathcal{Z}(\vec{J}) 
= \sum_{m=0}^{\infty} G^{(2m)} \frac{(\vec{J}^{2})^m}{(2m)!} = \sum_{m=0}^{\infty} G^{(2m)} \frac{\left( \sum_{i=1}^N J_i^2 \right)^{m}}{(2m)!} 
\nonumber \\
 & \ = \sum_{m=0}^{\infty} G^{(2m)} \frac{1}{(2m)!} \sum_{i_1 + \ldots + i_N = m} \! \! {m \choose i_1, i_2, \ldots, i_N} 
\prod_{k=1}^N J_{k}^{2i_{k}}
\label{eq:Zexpansion}
\end{align}
    and, by comparison with a generic Taylor expansion of $\mathcal{Z}(\vec{J})$, one finds
    \begin{equation}\label{eq:multiplicative_factor}
       G^{(n)}_{i_1 \ldots i_{n}} = G^{(n)} \frac{(n/2)!}{(n)!} \prod_{k=1}^N \frac{(n_k)!}{(n_k/2) !}\;,
    \end{equation}
    where $n_k$ is the number of indices $i$ which are equal to $k$. It should be noted that all Green functions or vertices with an odd number of indices of the same value
    vanish. From now on, we will only consider $G^{(n)} := G^{(n)}_{i \ldots i}$, as it contains all the relevant information. For instance, it directly follows from Eq.~(\ref{eq:multiplicative_factor}) that
    \begin{equation}
      \label{eq:G1122}
      G^{(4)}_{1122} = G^{(4)}_{1111} / 3 =  G^{(4)}/3 \;.
    \end{equation}
    Since all other generating functions share the $O(N)$ symmetry, they can also be expanded like $\mathcal{Z}(\vec{J})$ in Eq.~\eqref{eq:Zexpansion}.

  The relations between the different generating functions imply that their expansion coefficients are related. Taking derivatives of the above
  equations yields the following useful identities after some straightforward manipulations:
  \begin{align} 
     & G_c^{(0)}  = \ln \left(\frac{\mathcal{Z}}{\mathcal{Z}_0}\right) = -  \Gamma^{(0)}\;, 
    \label{eq:relations_gc0} \\
     & G_c^{(2)} = G^{(2)} = \left( \Gamma^{(2)} \right)^{-1}\;, 
    \label{eq:relations_gc2} \\
     & G_c^{(4)} = G^{(4)} - 3 \left(G^{(2)}\right)^2 = -\left( G^{(2)}_c \right)^{4} \Gamma^{(4)}\;.     \label{eq:relations_gc4}
  \end{align}
Similar identities can also be derived for higher-order correlation functions.

\section{Exact solution}
\label{sec:ExactSolution}

In order to calculate $\mathcal{Z} = \mathcal{Z} (0)$, we use hyperspherical coordinates to obtain
    \begin{equation} \label{eq:z_exact}
    	\mathcal{Z} = \int \mathcal{D}\vec{\varphi}\; \ee^{-S(\vec{\varphi})} =  \Omega_{N} R_{N-1}\;,
    \end{equation}
    where $\Omega_D$ is the surface area of the $D$-dimensional unit sphere and
    \begin{align} \label{eq:rk}
    	& R_k =  \int_0^{\infty} \dd x\; x^k\, \ee^{-\frac{r}{2}x^2 - \frac{u}{4!}x^4} 
= \: 2^{\frac{3k-5}{4}} 3^{\frac{k+1}{4}} u^{-\frac{k+3}{4}} 
\nonumber \\
 &  \quad \times
\Biggl[ \sqrt{u} \, \Gamma\left(\frac{k+1}{4}\right)  {}_1F_1 \left( \frac{k+1}{4};\frac{1}{2};\frac{3r^2}{2u} \right) \nonumber \\
    			 &  \qquad  {} -\sqrt{6} r \, \Gamma \left( \frac{k+3}{4} \right)  {}_1F_1 \left( \frac{k+3}{4};\frac{3}{2};\frac{3r^2}{2u} \right) \Biggr]\; ,
    \end{align}
where $ {}_1F_1 \left( a;b;z \right)$ is the Kummer confluent hypergeometric function.
       To determine $G^{(2)} = \langle\varphi_1^2\rangle$, it is again helpful to exploit symmetry arguments and use
$\langle\vec{\varphi}^2\rangle = \sum_{i=1}^N \langle\varphi_i^2 \rangle = N \langle\varphi_1^2 \rangle$ to arrive at
    \begin{equation} \label{eq:g2_exact}
    	G^{(2)} = \frac{1}{N} \langle \vec{\varphi}^2 \rangle = \frac{R_{N+1}}{N R_{N-1}} \;.
    \end{equation}
Making use of Eq.~\eqref{eq:G1122}, one easily calculates
    \begin{align}
    	\langle (\vec{\varphi}^2)^2 \rangle & =  N \langle \varphi_1^4 \rangle + N(N-1) \langle \varphi_1^2 \varphi_2^2 \rangle \nonumber \\
    	 & = N G^{(4)} + \frac{N(N-1)}{3} G^{(4)} \nonumber \\
    	 & = \frac{N(N+2)}{3} G^{(4)}\;,
    \end{align}
which results in
    \begin{equation} \label{eq:g4_exact}
    	G^{(4)} = \frac{3}{N(N+2)} \langle (\vec{\varphi}^2)^2 \rangle = \frac{3}{N(N+2)} \frac{R_{N+3}}{R_{N-1}}\;.
    \end{equation}    
    In principle, it is now also possible to go on to higher orders.

\begin{figure}[t]
    \centering
    \includegraphics[width=\columnwidth]{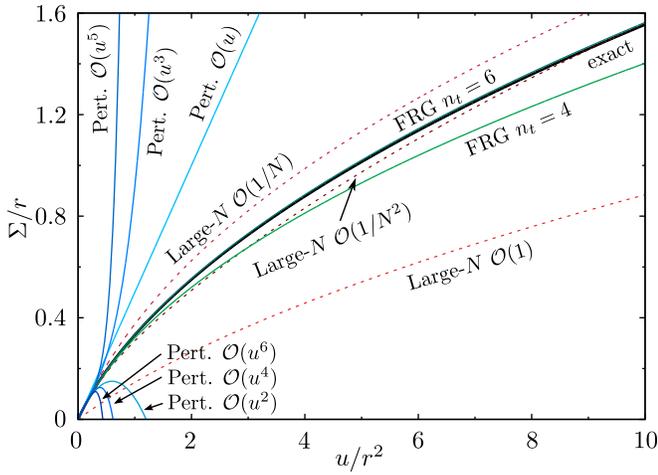}
    \caption{(colour online) Plots of $\Sigma$ for $N=1$. Within the resolution of this plot, the FRG results for $n_t =6$ lie on top of the exact results.}
    \label{fig:Gamma2_n1}
  \end{figure}

\begin{figure}
    \centering
    \includegraphics[width=\columnwidth]{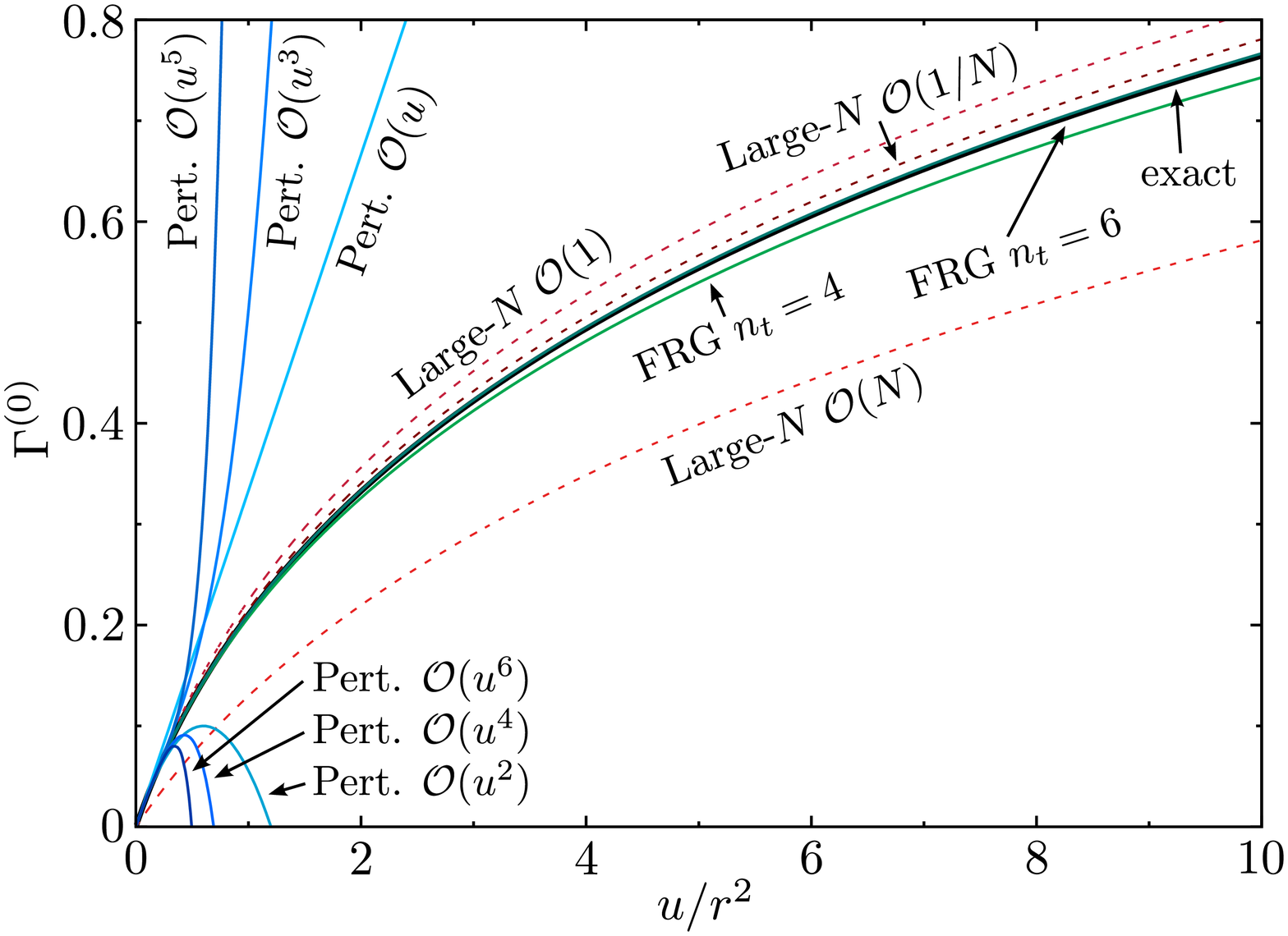}
    \caption{(colour online) Plots of $\Gamma^{(0)}$ for $N=2$. Within the resolution of this plot, the FRG results for $n_t =6$ lie on top of the exact results.}
    \label{fig:Gamma0_n2}
  \end{figure}

\begin{figure}[t]
    \centering
    \includegraphics[width=\columnwidth]{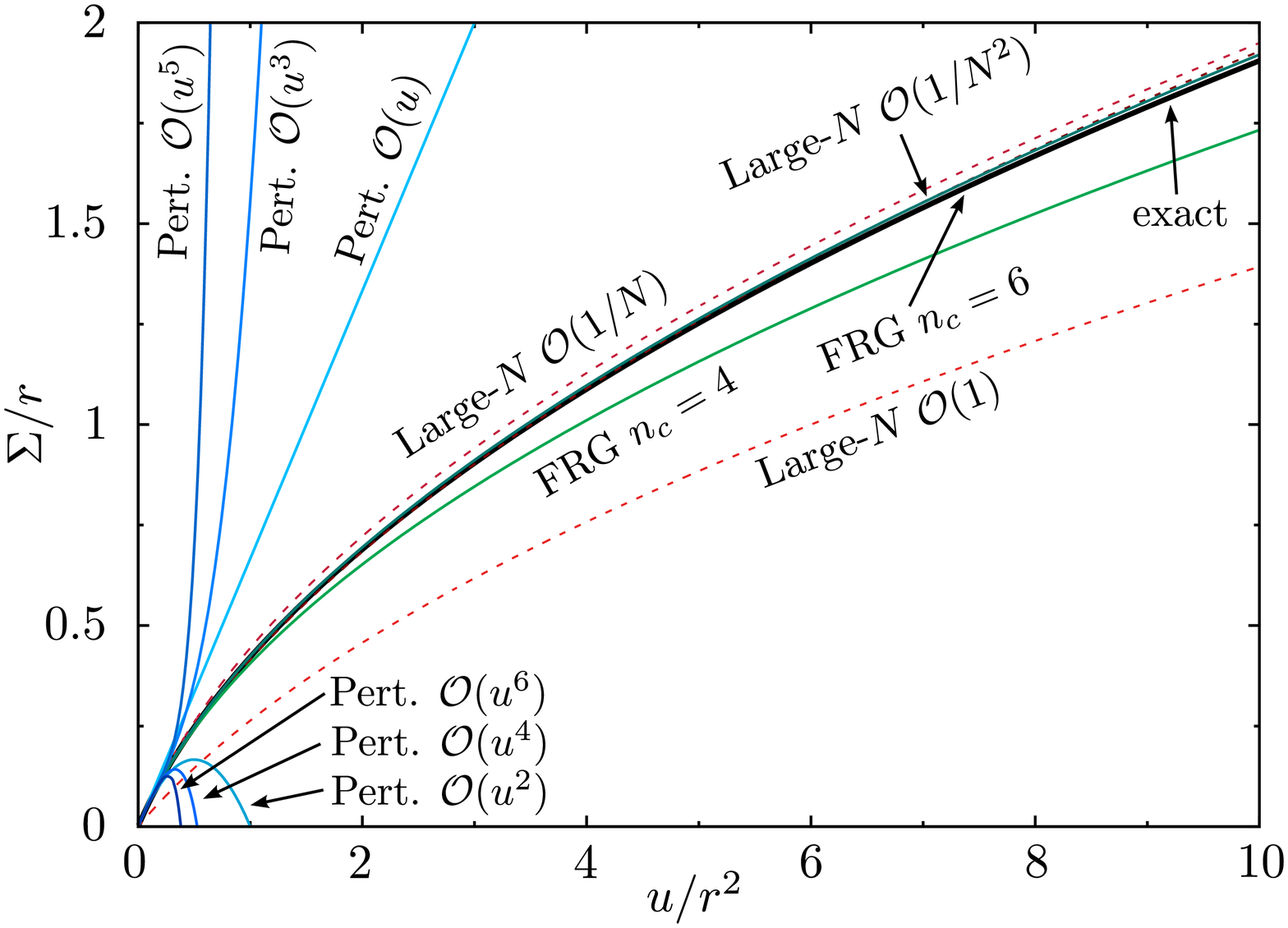}
    \caption{(colour online) Plots of $\Sigma$ for $N=2$. Within the resolution of this plot, the FRG results for $n_t =6$ lie on top of the exact results.}
    \label{fig:Gamma2_n2}
  \end{figure}

\begin{figure}[t]
    \centering
    \includegraphics[width=\columnwidth]{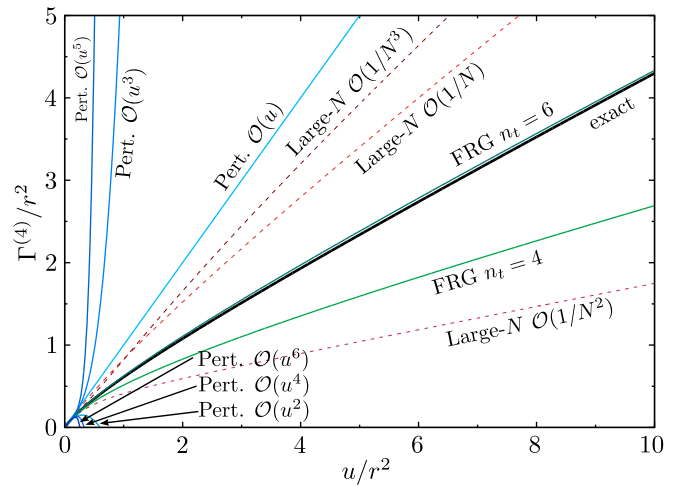}
    \caption{(colour online) Plots of $\Gamma^{(4)}$ for $N=2$. The FRG results for $n_t =6$ lie slightly above the exact results.}
    \label{fig:Gamma4_n2}
  \end{figure}

    Having found a method to compute the full Green functions $G^{(n)}$ in closed form, we can use Eqs.~\eqref{eq:relations_gc0}--\eqref{eq:relations_gc4} to calculate $G_c^{(n)}$ and $\Gamma^{(n)}$.
A plot of the self-energy $\Sigma = \Gamma^{(2)} - r$ for $N=1$ as well as plots of $\Gamma^{(0)}$, $\Sigma$, and $\Gamma^{(4)}$ for $N=2$ are shown in Figs.~\ref{fig:Gamma2_n1}--\ref{fig:Gamma4_n2} as a function of the dimensionless parameter $u/r^2$.
In these figures, we also show perturbative and non-perturbative results to be discussed in the following sections.


\section{Perturbation theory}
\label{sec:PerturbationTheory}

    As a next step, we will take a look at calculating Green functions within perturbation theory. The usual strategy is to expand the interaction term $\exp\left(-\frac{u}{4!}(\vec{\varphi}^2)^2\right)$
in powers of $u$. 
Introducing, in addition to the average with respect to the full action defined in Eq.~\eqref{eq:full_average},
the average with respect to the Gaussian part $\mathcal{S}_0$ of the action,
    \begin{equation} \label{eq:gaussian_average}
    	\langle \ldots \rangle_0 := \frac{\int \mathcal{D}\vec{\varphi}\,\ee^{-\mathcal{S}_0(\vec{\varphi})} \; [\ldots]}{\mathcal{Z}_0}\;,
    \end{equation}
the {\em connected} $n$-point functions are given by
    \begin{align} \label{eq:def_zj_perturbation}
    	G_c^{(n)} & = \left\langle \varphi_1^n \ee^{-\mathcal{S}_{\text{int}}(\vec{\varphi})}\right\rangle_0^{\text{conn}} \nonumber \\
        & = \sum_{k=0}^{\infty} \frac{1}{k!} \left\langle \varphi_1^n \left(- \frac{u}{4!} (\vec{\varphi}^2)^2 \right)^k \right\rangle_0^{\text{conn}} \;.
    \end{align}
The upper index ``conn'' indicates that only connected expressions should be considered when evaluating the Gaussian averages using Wick's theorem (see Refs.~\onlinecite{Negele88,Altland10,Zee10}). In fact, all occurring unconnected parts are cancelled by terms originating from an expansion of the partition function in the denominator of Eq.~\eqref{eq:full_average}, a result known as the {\em linked-cluster theorem}. 

In a nutshell, Wick's theorem states that $\langle \varphi_{i_1} \dots \varphi_{i_{2m}} \rangle_0$ equals all possible contractions of pairs of fields, e.g. $\langle \varphi_{i_1} \varphi_{i_2} \varphi_{i_3} \varphi_{i_4}  \rangle_0 = 
\langle \varphi_{i_1} \varphi_{i_2} \rangle_0 \langle \varphi_{i_3} \varphi_{i_4}  \rangle_0 + 
\langle \varphi_{i_1} \varphi_{i_3} \rangle_0 \langle \varphi_{i_2} \varphi_{i_4}  \rangle_0 + 
\langle \varphi_{i_1} \varphi_{i_4} \rangle_0 \langle \varphi_{i_2} \varphi_{i_3}  \rangle_0$ with $\langle \varphi_{i_1} \varphi_{i_2} \rangle_0 = \delta_{i_1,i_2} /r$. A large number of terms is generated and it is useful to identify the occurring terms with Feynman diagrams. 
In order to calculate the $\mathcal{O}(u^k)$ contribution
    	to $G^{(2m)}_c$, one can apply the following  {\em Feynman rules}:
    	\begin{enumerate}
    		\item Draw all topologically inequivalent connected diagrams $D_i$ 
with $k$ {\em wiggly} 4-point vertices \includegraphics{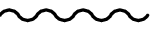}, 
and $2k+m$ propagators, which are denoted by a straight line \includegraphics{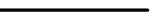}.
    		\item Assign an index $1 \leq j \leq N$ (or {\em colour}) to each line, keeping in mind that all external legs and
    				  both lines connected to each side of a vertex must have the same colour.
    		\item For each diagram, determine its {\em combinatorial factor} $C_i$, which is the number of Wick contractions it corresponds to (see below).
    		\item For each diagram, count the number of colours that can be chosen independently and denote it by $n_i$.
    		\item Assign a factor of ${1}/{r}$ to each propagator 
                  and a factor of ${-u}/{4!}$ to
    					each vertex. Multiply the result by $C_i N^{n_i}/k!$.
    		\item The $O(u^k)$ contribution to $G^{(2m)}$ is the sum of all these diagrams.
    	\end{enumerate}
    	
	\begin{figure}
	  \centering
    \includegraphics[width=\columnwidth]{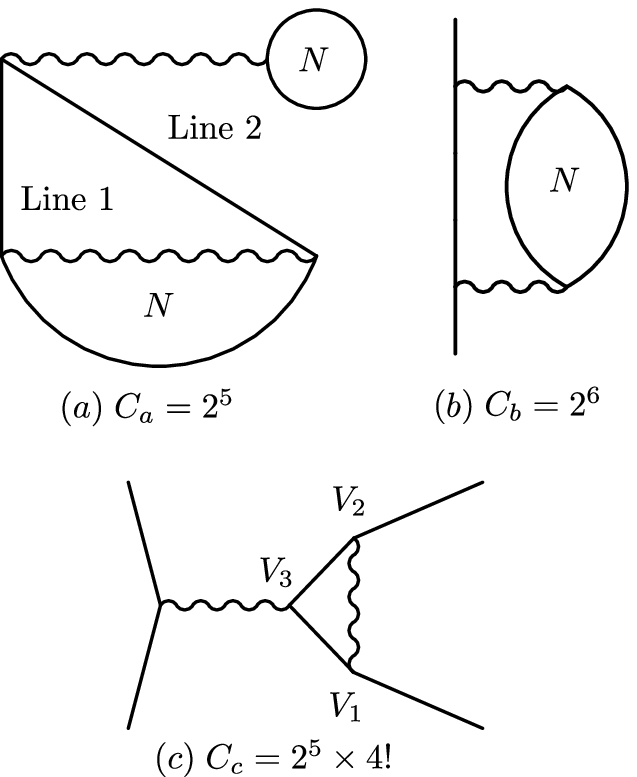}
	  \caption{Examples of Feynman diagrams. The straight and wiggly lines denote the free propagator $1/r$ and the interaction $u$, respectively.}
	  \label{fig:feynman_examples}
	\end{figure}
    	
    	Given our Feynman rules, all that is left to do is to apply them to determine $G_c^{(0)}$, $G_c^{(2)}$, and $G_c^{(4)}$. It
	is helpful to organise the diagrams by their number of vertices and number of independent colours.
        As many students have difficulties in figuring out the right combinatorial factors, let us discuss here the three examples depicted in Fig.~\ref{fig:feynman_examples}:
	\begin{enumerate}[(a)]
	 \item To determine $C_a$, consider the following reasoning: There are four locations to place the bubble, four ways of connecting Line 1 to the
	       second vertex and two more ways of connecting Line 2 to the second vertex. Having done that, the location of the remaining line is
	       already determined, hence $C_a = 4 \times 4 \times 2 = 2^5$.\\
	       As can be seen from the diagram, $n_c=2$ and therefore its total contribution is 
	       \begin{equation}
		 \frac{1}{2} \left(\frac{-u}{4!}\right)^2 \frac{1}{r^4} \times 2^5 \times N^2 = \frac{N^2}{36}\frac{u^2}{r^4}\;.
	       \end{equation}
	 \item There are eight places to connect the first external leg to and four places for the second leg. Since this already fixes the internal
	       line connecting the sides the legs are attached to, one has only an additional factor of two from arranging the lines that form the bubble
	       and thus $C_b = 8 \times 4 \times 2 =2^6$.
	       It follows that the diagram contributes
	      \begin{equation}
	        \frac{1}{2} \left(\frac{-u}{4!}\right)^2 \frac{1}{r^5} \times 2^6 \times N = \frac{N}{18}\frac{u^2}{r^5}\;.
	      \end{equation}	
	 \item One has four choices for attaching the two lines at $V_3$ to a vertex, four choices for the first line connecting $V_3$ to another vertex and two more choices how to connect $V_3$ to
	      $V_1$. Since there are exactly $4!$ ways to shuffle around the external legs, $C_c = 4 \times 4 \times 2 \times 4! = 2^5 \times 4!$.
	      Applying the remaining Feynman rules yields
	      \begin{equation}
		\frac{1}{2} \left(\frac{-u}{4!}\right)^2 \frac{1}{r^6} \times 2^5 \times 4! = \frac{2}{3}\frac{u^2}{r^6}\;.
	      \end{equation}

	\end{enumerate}

\begin{figure}
	  \centering
    \includegraphics[width=\columnwidth]{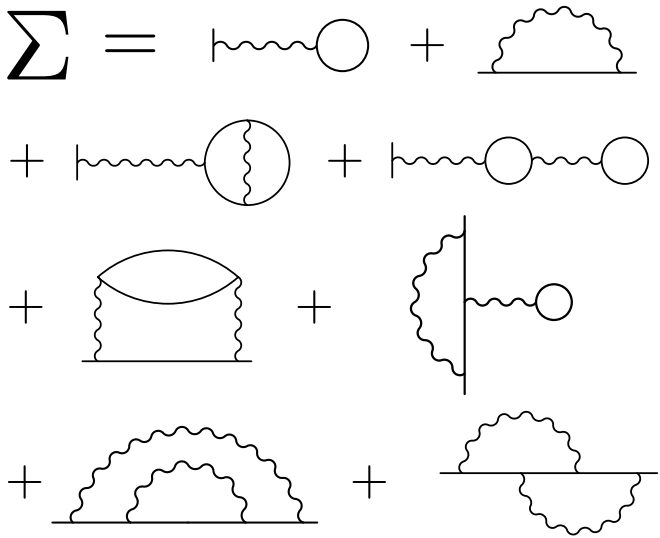}
	  \caption{Contributions to $\Sigma$ up to $\mathcal{O}(u^2)$.}
	  \label{fig:selfenergy}
	\end{figure}

	Adding up all the contributions for the connected Green functions leads to
	 \begin{align}
	   G_c^{(0)} = & \, - \frac{N^2 + 2N}{24} \frac{u}{r^2} + \frac{N^3+5N^2+6N}{144}\frac{u^2}{r^4} + \mathcal{O}\left(u^3\right) \;, \\
	   G_c^{(2)} = & \, \frac{1}{r} - \frac{N+2}{6} \frac{u}{r^3} + \frac{N^2+5N+6}{18} \frac{u^2}{r^5} + \mathcal{O}\left(u^3\right) \;, \\
	   G_c^{(4)} = & \, - \frac{u}{r^4} + \frac{5N+16}{6} \frac{u^2}{r^6} + \mathcal{O}\left(u^3\right) \;.
	 \end{align}
	With the connected Green functions determined, we can apply Eqs.~\eqref{eq:relations_gc0}--\eqref{eq:relations_gc4} to
	finally calculate the irreducible vertices. Alternatively, we obtain the irreducible vertices by simply ignoring the unconnected diagrams, stripping off the external legs, and finally multiplying the result by $-1$. For $n=2$, it is customary to subtract the inverse free propagator $r$ from $\Gamma^{(2)}$, leading to the self-energy $\Sigma = \Gamma^{(2)} -r$, whose graphical representation is shown in Fig.~\ref{fig:selfenergy}. 
Up to second order in $u$, we obtain
	  \begin{align}
	     \Gamma^{(0)} & = \frac{N^2+2N}{24} \frac{u}{r^2} - \frac{N^3+5N^2+6N}{144}\frac{u^2}{r^4} + \mathcal{O}\left(u^3\right) \;, 
\label{eq:Gamma0} \\
	     \Sigma & = \Gamma^{(2)} - r = \frac{N+2}{6} \frac{u}{r} - \frac{N^2+6N+8}{36} \frac{u^2}{r^3} + \mathcal{O}\left(u^3\right) \;, 
\label{eq:SigmaGamma2} \\
	     \Gamma^{(4)} & = u - \frac{N+8}{6} \frac{u^2}{r^2} + \mathcal{O}\left(u^3\right) \;. \label{eq:Gamma4}
             \end{align}
Comparing the perturbative results for $\Gamma^{(0)}$, $\Sigma$, and $\Gamma^{(4)}$ (as depicted in Figs.~\ref{fig:Gamma2_n1}--\ref{fig:Gamma4_n2}),
calculated within perturbation theory up to sixth order (see appendix), with the exact solution, we see that perturbation theory is only reliable for small $u/r^2$. 
As can be seen clearly in the figures, the usefulness of perturbation theory is restricted to small values of $u/r^2$ and going to higher order in perturbation theory does not help. This owes to the fact that our series expansions are not convergent at all and are only asymptotic series \cite{Schulman81,Negele88,Altland10}.
Roughly speaking, the vanishing radius of convergence is a natural consequence of the fact that all integrals above are actually divergent for $\textrm{Re}\,(u) < 0$. 
More rigorously, it is possible to deduce this vanishing radius of convergence from the series of expansion coefficients.
Even though the series are not convergent, truncating the series expansions at a finite given order can give rise to excellent results for very small values of $u/r^2$. In fact, for any given value of $u/r^2 < 3/(2(N-1))$, there is an optimal order 
\begin{equation}
n_{\text{max}} \sim \frac{1}{4} \left[\frac{3r^2}{u} - (N-1) + \sqrt{\frac{9r^4}{u^2} - \frac{6(N-1)r^2}{u}} \right] \;,
\end{equation}
beyond which an inclusion of higher-order terms gives rise to less accurate results.
While the residual error can be reduced for intermediate coupling strengths by using Borel summation techniques \cite{Negele88}, there clearly is a need for going beyond perturbation theory.

\section{Large-$N$ limit}
\label{sec:largeN}

The fact that $N$ is an intrinsically dimensionless parameter
      independent of any physical scale makes it especially suitable for taking limits \cite{Rossi98}. One expects that in the large-$N$ limit
      fluctuations are suppressed due to the infinite number of degrees of freedom to couple to and therefore some aspects of our theory might
      actually simplify.
To obtain a sensible limit,  one demands that all terms appearing in the action [see Eq.~\eqref{eq:action}] be extensive. Since $1/r$ is just the free propagator,
      it does not scale with $N$, thereby implying that $\vec{\varphi}^2 = \mathcal{O}(N)$. In order for $\mathcal{S}(\vec{\varphi})$ to be an extensive quantity, one thus finds that
      $u = \mathcal{O}(1/N)$ and it is reasonable to make the replacement $u \to {\ut}/{N}$. 

In the limit of large $N$, the leading and the next to leading order terms of the partition function $\mathcal{Z}$ can be calculated within the saddle point approximation \cite{Arfken05}.
Using hyperspherical coordinates and making the variable substitution $y = \vec{\varphi}^2/N$, the partition function turns into
\begin{equation}
  \label{eq:ZlargeN}
  \mathcal{Z} = \Omega_N N^{N/2} \int_0^{\infty} \frac{\mathrm{d}y}{2y} \mathrm{e}^{-N f(y)} \;,
\end{equation}
where the function $f(y)$ is defined by
\begin{equation}
  \label{eq:deff}
 f(y) = \frac{r}{2} y + \frac{\tilde u}{24} y^2 -\frac{1}{2} \ln(y) \;.
\end{equation}
This function has its minimum (for $r > 0$) at 
\begin{equation}
  \label{eq:y0}
  y_0 = \frac{3r}{\tilde u} \left( \sqrt{1 + \frac{2 \tilde u}{3 r^2}} - 1 \right) \;, 
\end{equation}
which in the limit $\tilde u \to 0$ is given by $y_0 = 1/r$.
Expanding $f(y)$ around this minimum up to second order in $y-y_0$ and performing the Gaussian integral we obtain
\begin{equation}
  \label{eq:ZlargeNtwo}
  \mathcal{Z} = \Omega_N N^{N/2} \left(\frac{2 \pi }{4 y_0^2 f^{\prime\prime}(y_0)}\right)^{1/2} \mathrm{e}^{-N f(y_0)} \left[1 + \mathcal{O}\left(\frac{1}{N} \right) \right] \;,
\end{equation}
with
\begin{equation}
  \label{eq:f0}
  f(y_0) = \frac{r y_0}{4} + \frac{1}{4} -\frac{1}{2} \ln\left(y_0 \right) \;,
\end{equation}
and
\begin{equation}
  \label{eq:f2}
  f^{\prime\prime}(y_0) = \frac{1}{y_0^2} - \frac{r}{2 y_0} \;,
\end{equation}
which for $\tilde u \to 0$ reduce to $f(y_0) = (1 + \ln r)/2$ and $f^{\prime\prime}(y_0) = r^2/2$.
We can now either use Stirling's formula to find an asymptotic expression for $\Omega_N N^{N/2}$ or notice that when  calculating the interaction-induced shift of the free energy, 
$\Gamma^{(0)} = - \ln \left(\frac{\mathcal{Z}}{\mathcal{Z}_0}\right)$, this term drops out.
It immediately follows that (see also Ref.~\onlinecite{Schelstraete94} for an alternative derivation)
\begin{align}
  \label{eq:Gamma0largeN}
  & \Gamma^{(0)} = N \left[ \frac{r y_0}{4} -\frac{1}{4} -\frac{1}{2} \ln\left(r y_0\right) \right] \nonumber \\
& \qquad \ \ {} + \frac{1}{2} \ln\left(2-r y_0\right) + \mathcal{O} \left(\frac{1}{N}\right) \;.
\end{align}
Higher-order terms in $1/N$ can be obtained by including fluctuation corrections to the saddle-point approximation (see appendix) and are generally much more difficult to compute. 

As for the two-point function $G_c^{(2)} = G^{(2)}$, using the saddle point approximation again, we have
\begin{equation}
  \label{eq:G2zero}
  G^{(2)} = \frac{\int_0^{\infty} \frac{\mathrm{d}y}{2y} \,y\,\mathrm{e}^{-N f(y)}}{\int_0^{\infty} \frac{\mathrm{d}y}{2y}\,\mathrm{e}^{-N f(y)}}
  = y_0 +\mathcal{O}\left(\frac{1}{N}\right) \;,
\end{equation}
i.e.\ $y_0$ is just the expectation value of $y$ in the limit $N \to \infty$.
Terms of arbitrary order in $1/N$ for any $n$-point function of the zero-dimensional $O(N)$ vector model can be obtained by taking successive derivatives of $\Gamma^{(0)}$ with respect to $r$ (see appendix).


To get a better understanding of which diagrams have actually been summed up, let us now study what the highest power of $N$ in a connected $n$-point function is. Assume that we are computing the $k$th order contribution to this Green function. There are exactly $k$ vertices, each
      of which has four points to connect a propagator to. Of those $4k$ points, only $4k - n$ remain unoccupied after attaching the external legs,
      which means that at most $2k - n/2$ colours can still be chosen independently of the colours of the external legs. Additionally, we are
      only interested in {\em connected} diagrams and one needs at least $k-1$ propagators to connect $k$ vertices. Since every propagator connecting
      two vertices further reduces the number of independent colours by one, we are left with $k + 1 - n/2$ colours that can be chosen freely, resulting
      in a contribution $\sim N^{k+1 - n/2}$. Taking into account the coupling constant ${\ut}/{N}$ of the vertices, one finally finds that
      the highest order contribution of $N$ is $N^{1 - n/2}$. Hence,
      \begin{equation} \label{eq:gc_vanish}
	G_c^{(n)} = \mathcal{O}\left(\frac{1}{N^{n/2-1}}\right) \;,
      \end{equation}
such that $\lim_{N \to \infty} G_c^{(n)} = 0$ for $n > 2$   
   while $G_c^{(2)}$ remains finite and $G_c^{(0)} = \mathcal{O}(N)$, as we have seen above.
 In the limit $N \to \infty$,  this implies that the disconnected Green functions factorise, i.e.\ for $i \neq j$ we have
      \begin{equation}
	\langle\varphi_i^2 \varphi_j^2 \rangle = \langle \varphi_i^2\rangle \langle\varphi_j^2\rangle \quad \mathrm{and} \quad \langle\varphi_i^4 \rangle = 3\left(\langle \varphi_i^2\rangle\right)^2 \;.
      \end{equation}
      As discussed in Ref.~\onlinecite{ZinnJustin02}, this is precisely the behaviour that one should expect, since the connected contributions to the correlation
      function can be identified as interaction-induced fluctuations which average out for large $N$.\\

      \begin{figure}	\centering
    \includegraphics[width=\columnwidth]{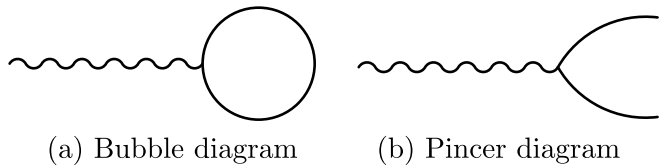}
	\caption{Building blocks of Feynman diagrams in the large-$N$ limit.}
	\label{fig:large_n_building_blocks}
      \end{figure}
We would now like to calculate $G_c^{(2)}$ diagrammatically in the large-$N$ limit, including terms of all orders in $\tilde u$. From the above consideration it is clear that
      only the diagrams with a maximum number of independent colours contribute, namely those that can be constructed from the two basic elements in
      Fig.~\ref{fig:large_n_building_blocks}. All other elements consist of pieces in which a propagator connects both ends of a vertex, thereby reducing
      the number of independent colours.\\

\begin{figure}
	\centering
    \includegraphics[width=\columnwidth]{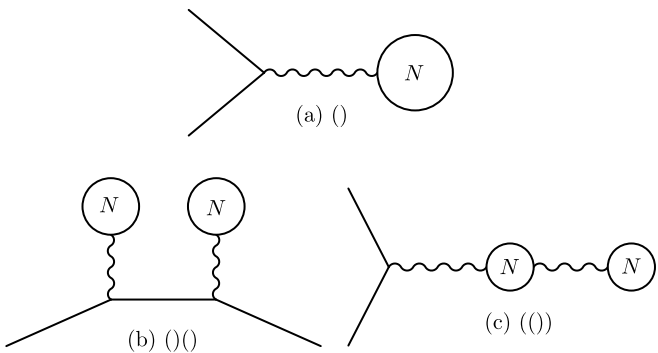}
	\caption{The first three diagrams and their short-hand notation.}
	\label{fig:large_n_shorthand}
      \end{figure}

      To determine all relevant diagrams, it is helpful to introduce shorthand notation using pairs of parentheses. One can  express the bubble
      [Fig.~\ref{fig:large_n_building_blocks} (a)] through an empty pair of parentheses ``$()$'',  where it is implied that two neighbouring bubbles
       ``$()()$'' are connected to each other. Furthermore, we denote by ``$(D)$'' a pincer [Fig.~\ref{fig:large_n_building_blocks} (b)], enclosing
      some arbitrary sub-diagram $D$. Examples of the notation are depicted in Fig.~\ref{fig:large_n_shorthand}.
      Fortunately, the combinatorial factor of diagrams constructed from those two elements only depends on the number of vertices and not on the internal
      structure. One thus finds the combinatorial factor of a diagram with $k$ vertices to be
      \begin{equation} \label{large_n_combinatorial_factor}
	{C}_k = 4^k \, k!\;.
      \end{equation}

      With $C_k$ determined, all that is left to do is count in how many ways one can arrange $k$ pairs of parentheses to give a valid expression. The answer
      to this well-known combinatorial problem is given by the {\em Catalan numbers}
      \begin{equation}
	\tilde C_k = \frac{(2k)!}{(k+1)! \, k!}\;.
      \end{equation}
      Putting everything together, we find that the $\ut^k$ contribution to the connected 2-point Green function is
      $\frac{1}{k! r} \left(- \frac{\tilde{u}}{4! r^2} \right)^k \tilde C_k \,C_k$ and that 
in the large-$N$ limit we have
      \begin{equation} \label{eq:g2c_large_n_all_orders}
	G_c^{(2)} = \frac{1}{r} \sum_{k=0}^\infty \frac{(2k)!}{k! \, (k+1)!} \left(-\frac{\tilde{u}}{6r^2}\right)^k \;,
      \end{equation}
which is in fact convergent for $|\ut| < {3r^2}/{2}$ and reproduces Eq.~\eqref{eq:G2zero}.



\begin{figure}
	\centering
    \includegraphics[width=\columnwidth]{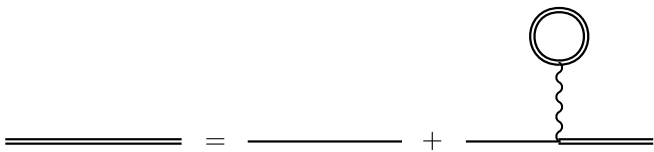}
	\caption{Self-consistent Hartree approximation of the full propagator. While the free propagator $1/r$ and the interaction $u = \tilde u/N$ are again denoted by a straight line and a wiggly line, the Hartree propagator $G_c^{(2)}$ is denoted by a double line.}
	\label{fig:hartree_equation}
      \end{figure}

\begin{figure}
	\centering
    \includegraphics[width=\columnwidth]{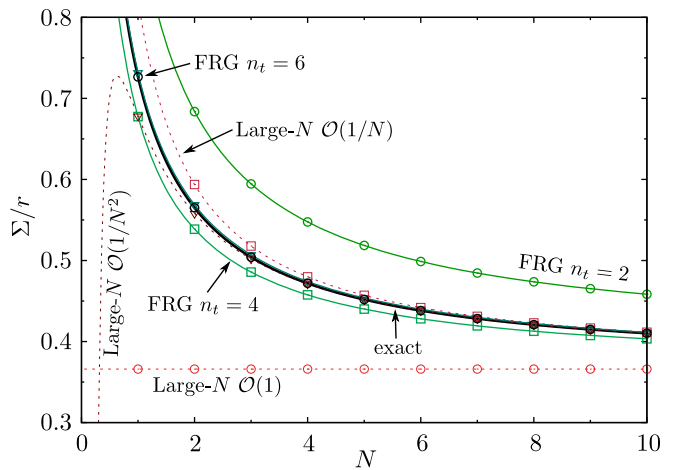}
	\caption{(colour online) Plot of the self-energy $\Sigma$ at fixed $\tilde u/r^2=3$ as a function of the continuous parameter $N$. Even though our model system has direct physical meaning only for integer $N$ (labelled by circles, triangles, and boxes), an evaluation of $\Sigma$ is possible in all cases also for non-integer values of $N$.}
	\label{fig:dependenceSigma_n}
      \end{figure}

\begin{figure}
	\centering
    \includegraphics[width=\columnwidth]{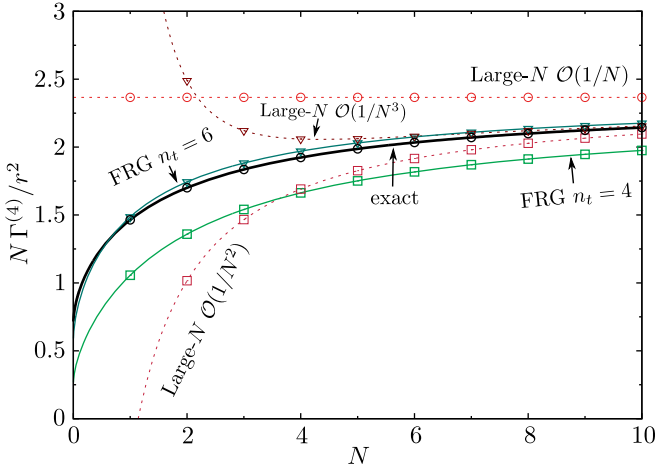}
	\caption{(colour online) Plot of the vertex $\tilde \Gamma^{(4)} = N \Gamma^{(4)}$ at fixed $\tilde u/r^2= N u/r^2 =3$ as a function of the continuous parameter $N$.}
	\label{fig:dependenceGamma4_n}
      \end{figure}

Examining the above contributions, it turns out that the large-$N$ expansion of the $N$-component field theory is identical to a {\em self-consistent Hartree approximation}.
      Recalling that the Bubble diagram above is also known as the {\em Hartree diagram}, one can write down a self-consistency equation for $G_c^{(2)}$,
      \begin{equation} \label{eq:hartree_selfconsistent}
	G_c^{(2)} = \frac{1}{r} - \frac{\tilde{u}}{3!} \frac{1}{r} \, \left( G_c^{(2)} \right)^2 \;,
      \end{equation}
      whose graphical counterpart is displayed in Fig.~\ref{fig:hartree_equation}.
Solving the above quadratic equation, we obtain
      \begin{equation} \label{eq:large_n_resummed}
	G_c^{(2)} = \frac{3r}{\tilde u} \left( \sqrt{1 + \frac{2 \tilde u}{3 r^2}} - 1 \right) = y_0 \;,
      \end{equation}
      which is the exact result for $G^{(2)} = G_c^{(2)}$ in the limit $N \to \infty$ and reproduces our previous results. 
By keeping all terms of order $\mathcal{O}(1/N)$ we also include the Fock term and many other diagrams. As a consequence,  $G^{(2)}$ then agrees for small $\tilde u/r^2$ with perturbation theory (and the exact result), but, in contrast to perturbation theory, it also gives accurate results even for small $N$ and larger values of $\tilde u/r^2$. This can be seen in  Figs.~\ref{fig:Gamma2_n1} and \ref{fig:Gamma2_n2}.
In Fig.~\ref{fig:dependenceSigma_n} we further plot $\Sigma$ for a fixed value of $\tilde u/r^2$ as a function of $N$, which we can assume to be a continuous parameter.
While $\Sigma$ and 
$\Gamma^{(0)}$ are quite well described by the large-$N$ expansion (if we include sufficiently many terms, see Figs.~\ref{fig:Gamma2_n1}--\ref{fig:Gamma2_n2}), the results for the vertex $\Gamma^{(4)}$ show considerable deviations from the exact result (see Fig.~\ref{fig:Gamma4_n2}).  Just as perturbation theory is an asymptotic expansion in the dimensionless parameter $u/r^2$, the large-$N$ expansion is an asymptotic expansion in the dimensionless parameter $1/N$. Although this asymptotic expansion is well behaved for both $\Gamma^{(0)}$ and $\Sigma$ up to $N = 1$, the asymptotic expansion for $\Gamma^{(4)}$ clearly breaks down for $N \approx 2$ (see Fig.~\ref{fig:dependenceGamma4_n}).

\section{FRG study}
\label{sec:FRG}

Finally, let us consider the effective average action FRG approach to our toy model.
An FRG study of the $N = 1$ zero-dimensional field theory
was proposed by Sch\"onhammer \cite{Schoenhammer01} and can be found in the lecture notes by Meden \cite{Meden03} or Pawlowski \cite{Pawlowski10} 
(see also Ref.~\cite{Salmhofer07} for a study of the case $N=2$). 
The same problem is dealt with in Exercise~7.1 of Ref.~\onlinecite{Kopietz10}, in which the reader is furthermore invited to show that due to its one-particle irreducibility the effective average action approach gives rise to much better results than the Wilson-Polchinski approach.

As has been mentioned in the introduction, in the effective average action approach a large artificial mass term (regulator) is initially added to all low-energy fluctuations. Replacing the action $\mathcal{S}$ by $\mathcal{S} + \Delta \mathcal{S}_\Lambda$,
a useful regulator term for the zero-dimensional $O(N)$ vector model is
\begin{equation}
  \label{eq:DeltaS}
  \Delta \mathcal{S}_\Lambda (\vec{\varphi}) = \frac{R_\Lambda}{2}  \vec{\varphi}^2 \;.
\end{equation}
This term effectively replaces the parameter $r$ in the action by $r + R_\Lambda$.
Subtracting a similar mass term from the flowing generating functional of one-particle irreducible vertices, 
  \begin{equation}
    \Gamma_\Lambda (\vec{\phi}) = \vec{J}_\Lambda (\vec{\phi}) \cdot \vec{\phi} - \gc \left(\vec{J}_\Lambda (\vec{\phi})\right) - \Delta \mathcal{S}_\Lambda (\vec{\phi}) \;,
  \end{equation}
this effective average action satisfies the initial condition 
\begin{equation} 
  \label{eq:flow_boundary}
  \lim_{R_\Lambda \to \infty} \Gamma_\Lambda (\vec{\phi}) = \mathcal{S} (\vec{\phi}) \;.
\end{equation}
Ultimately, we are interested in the problem without the artificial regulator term.
It is the aim of the FRG to follow the flow of various couplings as this regulator term is gradually removed. For $R_\Lambda \to 0$ we then obtain
\begin{equation} 
  \label{eq:flow_final}
  \lim_{R_\Lambda \to 0} \Gamma_\Lambda (\vec{\phi}) = \Gamma (\vec{\phi}) \;.
\end{equation}

As shown by Wetterich and others \cite{Wetterich93, Bonini93, Morris94}, the effective average action obeys the following flow equation,
\begin{equation}
  \label{eq:flowequation}
        \partial_\Lambda \Gamma_\Lambda (\vec{\phi}) = 
        \frac{1}{2} \Tr \left[ [\pla \mathbf{R}_\Lambda] \left( \pd{}{\vec{\phi}} \otimes \pd{}{\vec{\phi}}\:\gwl(\vec{\phi})
	+ \mathbf{R}_\Lambda \right)^{-1} \right] \;,
\end{equation}
where $\mathbf{R}_\Lambda$ is just $R_\Lambda$ times the unit matrix and 
\begin{equation}
  \label{eq:Gammamatrixelements}
  \left( \pd{}{\vec{\phi}} \otimes \pd{}{\vec{\phi}}\:\gwl(\vec{\phi}) \right)_{ij}
  = \partial_{\phi_i} \partial_{\phi_j}
  \Gamma_\Lambda  (\vec{\phi}) \;.
\end{equation}
If we are also interested in the interaction-induced shift of the free energy, $\Gamma^{(0)}$, we have to subtract from the r.h.s.\ of Eq.~\eqref{eq:flowequation} the same term with $u$ set equal to zero.

\begin{figure}
    \centering
    \includegraphics[width=\columnwidth]{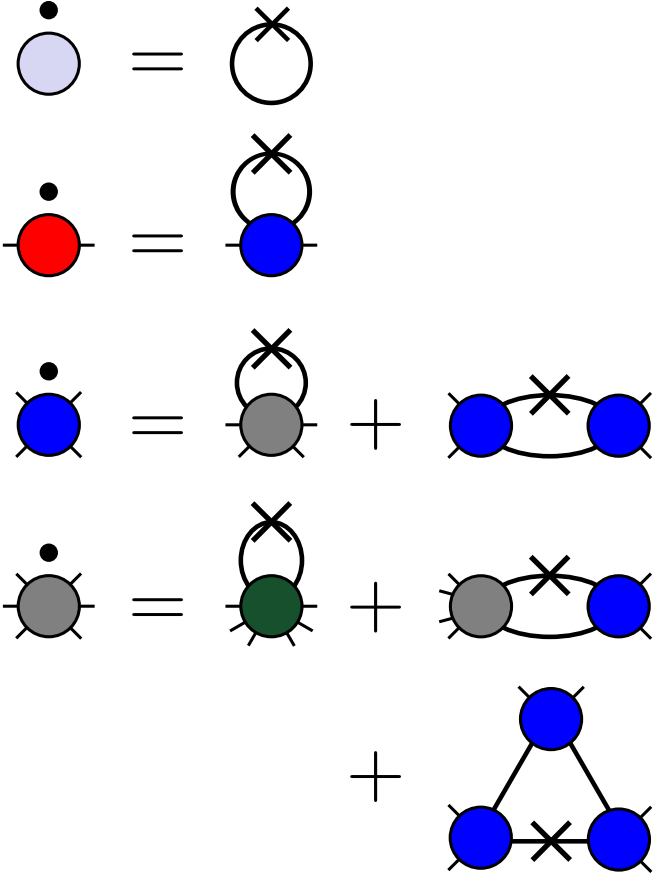}
    \caption{(colour online) Graphical representation of the flow equations for the first four running couplings. While the $n$-point vertices $\Gamma_\Lambda^{(n)}$ are denoted by filled circles with $n$ external legs and the derivative with respect to $\Lambda$ is denoted by a dot, the flowing propagator $G_\Lambda ^{(2)} = \left(\Gamma_\Lambda^{(2)} + R_\Lambda \right)^{-1}$ is represented by a thick line. The cross appearing exactly once in every graph on the r.h.s.\ simply denotes the derivative of the regulator $R_\Lambda$ with respect to $\Lambda$.}
    \label{fig:flow_equations}
   \end{figure}

  One way of solving Eq.~\eqref{eq:flowequation} is to expand both sides of this equation with respect to the fields and compare their coefficients to derive flow equations for the expansion
  coefficients of $\gal (\vec{\phi})$. This is the so-called {\em vertex expansion}.
Proceeding along this way, one obtains a system of infinitely many coupled ordinary differential equations, of which we list the first four,
  \begin{align}
&    \pla \Gamma^{(0)}_{\Lambda} = \frac{N}{2} \left( \partial_\Lambda R_\Lambda \right) 
\left( G_\Lambda - G_{0,\Lambda} \right) \;, \\
&    \pla \Gamma_{\Lambda}^{(2)} = - \frac{N+2}{6} \, \Gamma^{(4)}_{\Lambda} \left( \partial_\Lambda R_\Lambda \right) G_\Lambda^2 \;, \\
&    \pla \Gamma_\Lambda^{(4)} = - \frac{N+4}{10} \, \Gamma^{(6)}_{\Lambda} \left( \partial_\Lambda R_\Lambda \right) G_\Lambda^2 \nonumber \\ 
& \qquad \qquad {} + \frac{N+8}{3} \, \left(\Gamma^{(4)}_{\Lambda}\right)^2 \left( \partial_\Lambda R_\Lambda \right)  G_\Lambda^3 \;, \\
&  \pla \Gamma^{(6)}_{\Lambda} = -  
\frac{N+6}{14} \, \Gamma_\Lambda^{(8)} \left( \partial_\Lambda R_\Lambda \right) G_\Lambda^2 \nonumber \\
& \qquad \qquad {} 	+ (N+14) \, \Gamma^{(6)}_{\Lambda} \Gamma^{(4)}_{\Lambda} \left( \partial_\Lambda R_\Lambda \right) G_\Lambda^3 \nonumber \\
& \qquad \qquad {} 	-  \frac{5N+130}{3} \,
\left(\Gamma_{\Lambda}^{(4)}\right)^3 \left( \partial_\Lambda R_\Lambda \right) G_\Lambda^4 \;.
  \end{align}
Here, $G_\Lambda = 1/(\Gamma_\Lambda^{(2)} + R_\Lambda)$ is the flowing propagator and $G_{0,\Lambda} = 1/(r + R_\Lambda)$ is the corresponding non-interacting propagator. Graphical representations of these flow equations are depicted in Fig.~\ref{fig:flow_equations}.


We {\em truncate} our vertex expansion by setting all vertices higher than a given order $n_t$ equal to their
    initial values. Choosing $n_t \in \{2,4,6\}$, the resulting system of ordinary differential equations consists of very few equations and can easily be solved numerically.
As the couplings $\Gamma^{(n)}$ with $n \geq 2$ are completely decoupled from $\Gamma^{(0)}$, we do not need to keep track of the latter if we are not interested in the flow of the interaction-induced shift of the free energy. Even when truncating the system of flow equations at $n_t = 6$ we therefore only need to consider three flowing couplings.
    According to Eq.~\eqref{eq:flow_boundary}, our initial values for the $\Gamma_{\Lambda_0}^{(n)}$ are
    \begin{align}
      \Gamma_{\Lambda_0}^{(0)} & = 0 \;,\quad \Gamma_{\Lambda_0}^{(2)} = r  \;, \nonumber \\
      \Gamma_{\Lambda_0}^{(4)} & = u \,, \quad \Gamma_{\Lambda_0}^{(n)} = 0 \quad \textrm{for} \quad n \geq 6 \;.\label{eq:flow_boundary_evaluated}
    \end{align}
To numerically integrate our set of flow equations, we use the following regulator,
\begin{equation}
  \label{eq:regulator}
  R_\Lambda = \Lambda^{-1} - r \;,
\end{equation}
which leads to $\partial_\Lambda R_\Lambda = -1/\Lambda^2$, $G_{0,\Lambda} = \Lambda$, and $G_\Lambda = 1/(\Gamma_\Lambda^{(2)} + \Lambda^{-1} - r)$. 
For $N=1$, the resulting equations agree with Refs.~\onlinecite{Meden03,Kopietz10}.
Since $R_\Lambda$ diverges for $\Lambda \to 0$ and vanishes for $\Lambda \to r^{-1}$, our flow equations have to be integrated from $\Lambda = 0$ to $\Lambda = r^{-1}$. 
As can be seen in Figs.~\ref{fig:Gamma2_n1}--\ref{fig:Gamma4_n2}, the values obtained by setting $n_t = 6$ already turn out to be very close
    to the exact result and thereby reduce the possible benefit that could be gained by including higher orders.

    \begin{figure}
      \centering
    \includegraphics[width=\columnwidth]{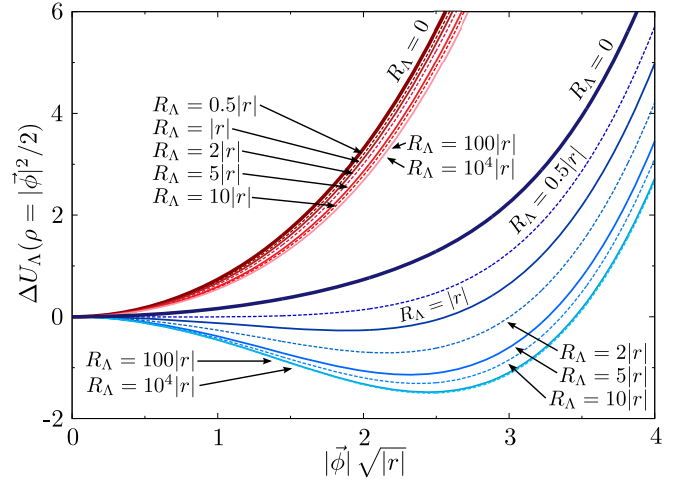}
      \caption{(colour online) Evolution of $\Delta U_\Lambda(\rho=\vec{\phi}^2/2)$ for $u/r^2 = 1$ and $r > 0$ [upper (red) curves] as well as $u/r^2 = 1$ and $r < 0$ [lower (blue) curves]. It should be observed that the final (physical) curve with $R_\Lambda = 0$ always has a single-valued minimum located at $\rho = 0$, owing to the fact that there is no spontaneous symmetry breaking in a system of finite (or even zero) volume.}
      \label{fig:gamma_full_both}
    \end{figure}
So far, we have restricted our discussion of the zero-dimensional $O(N)$ vector model to the case of positive $r$. While the perturbative approach is only meaningful for such values of $r$, the large-$N$ expansion and the FRG study can readily be generalised to negative $r$. For the large-$N$ expansion we note that for general $r$ the minimum of $f(y)$ is located at 
$y_0 = \frac{3r}{\tilde u} \left( \textrm{sgn} (r) \sqrt{1 + \frac{2 \tilde u}{3 r^2}} - 1 \right)$, which for negative $r$ leads to slight modifications of all equations.
As concerns the (FRG) vertex expansion, the extension to negative $r$ is possible by expanding the effective average action $\Gamma_\Lambda (\vec{\phi}) $ around its flowing minimum. 

Besides the vertex expansion, the derivative expansion is a powerful approximation for the FRG. Within this approximation, terms of arbitrary order are kept for a homogeneous field configuration. As there are no spatial variations in a zero-dimensional field theory, the derivative approximation actually becomes an exact method to solve the zero-dimensional $O(N)$ vector model.
Because $\Gamma_{\Lambda}(\vec{\phi})$ is only a function of $|\vec{\phi}|$, it is customary to define $\rho = {\vec{\phi}^2}/{2} = \sum_{i=1}^N \phi_i^2/2$ and introduce the flowing effective potential as $U_\Lambda (\rho) = \Gamma_\Lambda (\vec{\phi})$.
Using
\begin{equation}
  \label{eq:GammaU}
       \frac{\partial^2 U_\Lambda (\rho)}{\partial \phi_i \partial \phi_j} = 
       U_\Lambda^\prime(\rho) \delta_{i,j} + U_\Lambda^{\prime\prime} (\rho)  \phi_i \phi_j \,,
\end{equation}
and letting $\vec{\phi}$ point along one of the coordinate axes, we have
    \begin{align}
     & \pla U_\Lambda (\rho)  = \frac{1}{2} \Tr \left[ [\pla \mathbf{R}_\Lambda] \left( \pd{}{\vec{\phi}} \otimes \pd{}{\vec{\phi}}\:U_\Lambda(\rho)
	+ \mathbf{R}_\Lambda \right)^{-1} \right] \nonumber \\
      &\quad = \frac{\pla R_{\Lambda}}{2} \sum_{i=1}^N \left( \frac{\partial^2 U_\Lambda}{\partial \phi_i^2}
	+ R_{\Lambda} \right)^{-1} \nonumber \\
      & \quad = \frac{\pla R_{\Lambda}}{2} \left( \frac{1}{U_\Lambda^\prime(\rho) + 2 \rho U_\Lambda^{\prime\prime} (\rho)  + R_{\Lambda}}  + \frac{N-1}{U_\Lambda^\prime(\rho)  + R_{\Lambda}} \right) \;. \label{flow_eq_wetterich_simplified}
    \end{align}
    Defining $U_\Lambda^{(n)}$ as the expansion coefficients of $U_\Lambda (\rho)$ as in
    \begin{equation} \label{def_lambda_tilde}
      U_\Lambda (\rho) = \sum_{n=0}^{\infty} \frac{U_{\Lambda}^{(n)}}{n!} \rho^n\;,
    \end{equation}
    one can compare the series of $U_\Lambda (\rho)$ and $\gwl(|\vec{\phi}|)$,
    \begin{equation}
      \gal(|\vec{\phi}|) 
= \sum_{n=0}^{N} \frac{\gal^{(2n)}}{(2n)!}
      (2\rho)^n =  \sum_{n=0}^{\infty} \frac{U_{\Lambda}^{(n)}}{n!} \rho^n \;,
    \end{equation}
    to find that
    \begin{equation} \label{relation_iv_wetterich_iv}
      \gal^{(2n)} = (2n-1)!! \, U_{\Lambda}^{(n)}\;.
    \end{equation}
Solving the partial differential equation \eqref{flow_eq_wetterich_simplified} for $U_\Lambda (\rho)$ by discretising $\rho$, we obtain results which are very close to the exact result.
We can then use Eq.~\eqref{relation_iv_wetterich_iv} to obtain the vertex functions that we were initially interested in.
In Fig.~\ref{fig:gamma_full_both} we show the evolution of the effective potential $U_\Lambda (\rho)$ for both a positive and a negative value of $r$. 
This evolution is especially interesting for negative $r$.
At the beginning of the FRG flow, all fluctuations are excluded, such that the minimum of the effective average potential is located at a finite value of $\rho$. Neglecting fluctuations would therefore lead (as in mean-field theory) to a symmetry-broken state by picking one vector $\vec{\phi}$ which satisfies $\rho = \vec{\phi}^2/2$. 
But we do have to include fluctuations, and this is what the FRG does. 
As all fluctuations are integrated out, the minimum of $U_\Lambda(\rho)$ is moved to $\rho = 0$, giving rise to a single, non-degenerate ground state with $\vec{\phi} = 0$.
This result is a natural consequence of the fact that there is no spontaneous symmetry breaking in any system of finite (including zero) size. Let us also note that the effective potential becomes a convex function of $\vec{\phi}$, as one should expect for a function whose origin lies in a Legendre transformation.
It is impossible to recover this feature within the vertex expansion.


\section{Conclusions}
  \label{sec:Conclusion}
  
  In order to conclude our comparison of different approaches to the zero-dimensional $O(N)$ vector model, let us come back to the plots in Figs.~\ref{fig:Gamma2_n1}--\ref{fig:Gamma4_n2}.
The most obvious fact to observe is that perturbation theory consistently breaks down for $u/r^2 = \mathcal{O} (1)$ (or, more accurately, for $\tilde u/r^2 = \mathcal{O} (1)$ if  $N \geq 1$). 
Furthermore, it can be seen that including higher orders of the expansion parameter $u$ only exacerbates the situation in the strong coupling regime---exactly the behaviour one would expect from an asymptotic series whose contributions have alternating signs.

  Moving on, we see that when including the first two subleading orders of the self-energy $\Sigma$, the large-$N$ expansion makes surprisingly good predictions for $N=2$ and even $N=1$. 
However, the large-$N$ expansion fails for the calculation of the four-point vertex $\Gamma^{(4)}$ for both $N=1$ and $N=2$, and one needs to go to much larger values of $N$ (or very small values of $\tilde u/r^2$) to see an asymptotic convergence.
As concerns more involved model systems, terms of order $\mathcal{O}(1/N^2)$ are usually very difficult to compute, and it is widely known that the large-$N$ expansion can fail, as it does for instance for half-odd integer spin antiferromagnets \cite{Auerbach94}.

  Last of all, let us come back to the FRG results. Within the vertex expansion, one truncates the system of flow equations at a given order and including more terms in this expansion consistently improves this result. As can be seen in Figs.~\ref{fig:Gamma0_n2}--\ref{fig:Gamma4_n2}, for the results with $n_t = 6$ (keeping all vertices up to the six-point vertex) 
 it becomes difficult to distinguish the FRG results from the exact values no matter how large the interaction strength $u/r^2$ is.
As higher-order vertices are more directly affected by our truncations, the results decrease in their precision when going to higher-order vertices. 
Despite of that, one can clearly see that the FRG outperforms the other approaches with only three running couplings.
For the zero-dimensional $O(N)$ vector model considered here, the FRG flow equation can directly be solved using numerical methods. This is not possible in higher dimensions, as the space of possible functions $\vec{\varphi}(x)$ becomes just too big \cite{Hedden04}.
However, keeping in a derivative expansion in addition to the effective potential $U_\Lambda(\rho)$ just the derivative term as already present in the action of a $\varphi^4$ theory leads to a very useful approximation, the local potential approximation (LPA) \cite{Berges02,Delamotte07,Kopietz10}. It then turns out, that while the effective action is still convex, its minimum is not single-valued. Instead, one obtains a flat plateau, giving rise to spontaneous symmetry breaking.

  Summing up, we find that perturbation theory only works for small values of $u/r^2$. On the other hand, both the large-$N$ expansion and the functional renormalisation group can be used to explore non-perturbative aspects of our field theory. However, as the large-$N$ expansion fails to determine the four-point vertex for $N =1$ or $2$, the FRG turns out to be most reliable for small $N$.

\acknowledgments

We thank Holger Gies, Aldo Isidori and Peter Kopietz for valuable discussions. 
This work was supported by the German Research Foundation DFG
through the research group FOR\,723.

\appendix
\section{Higher-order perturbation theory and results for the large-$N$ expansion}

In this appendix we collect some results for perturbation theory up to sixth order and also give the next-to-next leading order results of the one-particle irreducible $n$-point vertices in the large-$N$ expansion for $n \leq 4$. These expressions are used in Figs.~\ref{fig:Gamma2_n1}--\ref{fig:Gamma4_n2} as well as Figs.~\ref{fig:dependenceSigma_n} and \ref{fig:dependenceGamma4_n}.

Generalising the calculations leading to Eqs.~(\ref{eq:Gamma0})--(\ref{eq:Gamma4}) to higher order in $u$ using Mathematica, we obtain
  \begin{widetext}
  \begin{align}
    \Gamma^{(0)} =& \frac{N^2 + 2N}{24} \frac{u}{r^2} - \frac{N^3+5N^2+6N}{144} \left(\frac{u}{r^2}\right)^2 + \frac{5N^4+44N^3+128N^2+120N}{2592} \left(\frac{u}{r^2}\right)^3 \nonumber \\ &- \frac{7N^5+93N^4+468N^3+1040N^2+840N}{10368}\left(\frac{u}{r^2}\right)^4 \nonumber\\ &+ \frac{21N^6 + 386N^5 + 2900 N^4 + 11000N^3 + 20712 N^2 + 15120N}{77760} \left(\frac{u}{r^2}\right)^5 \nonumber \\ &- \frac{33N^7 + 793N^6 + 8178 N^5 + 45900 N^4 + 146000 N^3 + 245352 N^2 + 166320 N}{279936}\left(\frac{u}{r^2}\right)^6 + \mathcal{O}\left(\left(\frac{u}{r^2}\right)^7\right) \;,
  \end{align}
  \begin{align}
    \frac{\Sigma}{r} =& \frac{N^2 + 2N}{6} \frac{u}{r^2} - \frac{N^2+6N+8}{36} \left(\frac{u}{r^2}\right)^2 + \frac{N^3+11N^2+38N+40}{108}\left(\frac{u}{r^2}\right)^3 \nonumber\\ &- \frac{5N^4+84N^3+512N^2+1320N+1184}{1296}\left(\frac{u}{r^2}\right)^4 \nonumber \\
 & {} + \frac{7N^5+163N^4+1492N^3+6640N^2+14152N+11296}{3888}\left(\frac{u}{r^2}\right)^5 \nonumber \\ &- \frac{21N^6+638N^5+8020N^4+53000N^3+192232N^2+357680N+261184}{23328} \left(\frac{u}{r^2}\right)^6+ \mathcal{O}\left(\left(\frac{u}{r^2}\right)^7\right) \;,
  \end{align}
  \begin{align}
    \frac{\Gamma^{(4)}}{r^2} =& \frac{u}{r^2} - \frac{N+8}{6}\left(\frac{u}{r^2}\right)^2 + \frac{3N^2+46N+140}{36}\left(\frac{u}{r^2}\right)^3 - \frac{5N^3+117N^2+772N+1536}{108} \left(\frac{u}{r^2}\right)^4 \nonumber \\ &+ \frac{35N^4+1124N^3+11880N^2+51568N+79168}{1296}\left(\frac{u}{r^2}\right)^5 \nonumber \\ &- \frac{63N^5+2609N^4+38874N^3+271676N^2+906576N+1164032}{3888}\left(\frac{u}{r^2}\right)^6 + \mathcal{O}\left(\left(\frac{u}{r^2}\right)^7\right) \;.
  \end{align}
  \end{widetext}

To obtain the next-to-next leading order term in the large-$N$ expansion of $\Gamma$, we can expand $f(y)$ in the partition function Eq.~\eqref{eq:ZlargeN} around its minimum up to fourth order in $y-y_0$ and also expand the prefactor $1/y$ up to second order in $y-y_0$. Collecting all the relevant terms, Eq.~\eqref{eq:ZlargeNtwo} gets replaced by
\begin{align}
  \label{eq:ZlargeN4}
  \mathcal{Z} = & \Omega_N N^{N/2} \left(\frac{2 \pi }{4 y_0^2 f^{\prime\prime}(y_0)}\right)^{1/2} \mathrm{e}^{-N f(y_0)} \nonumber \\
 {} & \times \left[ 1 + \frac{12 r^2 y_0^2 - 27 r y_0 + 16}{6 N (2-r y_0)^3} \right] \left[1 + \mathcal{O}\left(\frac{1}{N} \right) \right] \;.
\end{align}
Dividing $\mathcal{Z}$ by $\mathcal{Z}_0$ and taking the logarithm we obtain for the interaction-induced free energy shift
\begin{align}
  \label{eq:Gamma0largeNappendix}
  & \Gamma^{(0)} = N \left[ \frac{r y_0}{4} -\frac{1}{4} -\frac{1}{2} \ln\left(r y_0\right) \right]  + \frac{1}{2} \ln\left(2-r y_0\right) \nonumber \\
& \qquad \ \ {} - \frac{(8+r y_0)(r y_0 -1)^2}{6 N (2-r y_0)^3}
+ \mathcal{O} \left(\frac{1}{N^2}\right) \;.
\end{align}
While higher-order terms in $1/N$ can be evaluated by generalising the above approach, a simple identity between various derivatives of the free energy allows for a recursive evaluation of higher-order terms \cite{Schelstraete94}.
Taking derivatives of $\Gamma^{(0)}$ with respect to $r$, we furthermore obtain 
\begin{align}
  & \Sigma = \frac{1}{y_0}-r  + \frac{2 (1 - r y_0)}{N y_0 (2 - r
   y_0)^2} \nonumber \\
 & \ \ {} + \frac{4 (r y_0 - 1)^2 (3 r y_0 - 1)}{N^2 y_0 (2 - r y_0)^5} 
+ \mathcal{O} \left(\frac{1}{N^3}\right) \;,
  \end{align}
  \begin{align}
  & \Gamma^{(4)} = \frac{6 (1 - r y_0)}{N y_0^2 (2-r y_0)}
-\frac{12 (1 - r y_0)^2 (r^2 y_0^2 - 3 r y_0 + 6)}{N^2
   y_0^2 (2 - r y_0)^4} \nonumber \\
& \ \ {} + \frac{24 (1 - r y_0)^3 (r^4 y_0^4-8 r^3 y_0^3+35 r^2 y_0^2-49 r
   y_0+56)}{N^3 y_0^2 (2 - r y_0)^7} \nonumber \\
& \ \ {} + \mathcal{O} \left(\frac{1}{N^4}\right) \;.
\end{align}




\begin{thebibliography}{10}
\expandafter\ifx\csname url\endcsname\relax
  \def\url#1{{\tt #1}}\fi
\expandafter\ifx\csname urlprefix\endcsname\relax\def\urlprefix{URL }\fi
\providecommand{\eprint}[2][]{\url{#2}}

\bibitem{Schulman81}
Schulman L~S 1981 {\em Techniques and applications of path integrals\/} (Wiley,
  New York)

\bibitem{Negele88}
Negele J~W and Orland H 1988 {\em {Quantum many-particle systems}\/}
  (Addison-Wesley, Redwood City)

\bibitem{Altland10}
Altland A and Simons B 2010 {\em {Condensed Matter Field Theory}\/} 2nd ed
  (Cambridge University Press, Cambridge, England)

\bibitem{Zee10}
Zee A 2010 {\em {Quantum Field Theory in a Nutshell}\/} 2nd ed (Princeton
  University Press, Princeton, New Jersey)

\bibitem{Schelstraete94}
Schelstraete S and Verschelde H 1994 {\em Phys. Lett. B\/} {\bf 332} 36--43

\bibitem{Polchinski84}
Polchinski J 1984 {\em Nucl. Phys. B\/} {\bf 231} 269--295

\bibitem{Wetterich93}
Wetterich C 1993 {\em Phys. Lett. B\/} {\bf 301} 90--94

\bibitem{Bonini93}
Bonini M, D'Attanasio M and Marchesini G 1993 {\em Nucl. Phys. B\/} {\bf 409}
  441--464

\bibitem{Morris94}
Morris T~R 1994 {\em Int. J. Mod. Phys. A\/} {\bf 9} 2411--2449

\bibitem{Keller91}
Keller G and Kopper C 1991 {\em Phys. Lett. B\/} {\bf 273} 323--332

\bibitem{Morris98}
Morris T~R 1998 {\em Progr. Theoret. Phys. Suppl.\/} {\bf 131} 395--414

\bibitem{Bagnuls01}
Bagnuls C and Bervillier C 2001 {\em Phys. Rep.\/} {\bf 348} 91--157

\bibitem{Berges02}
Berges J, Tetradis N and Wetterich C 2002 {\em Phys. Rep.\/} {\bf 363} 223--386

\bibitem{Gies06}
Gies H 2006 {Introduction to the functional RG and applications to gauge
  theories} (\textit{Preprint} \eprint{arXiv:hep-ph/0611146})

\bibitem{Pawlowski07}
Pawlowski J~M 2007 {\em Ann. Phys.\/} {\bf 322} 2831--2915

\bibitem{Delamotte07}
Delamotte B 2007  (\textit{Preprint} \eprint{arXiv:cond-mat/0702365})

\bibitem{Kopietz10}
Kopietz P, Bartosch L and Sch\"{u}tz F 2010 {\em {Introduction to the
  Functional Renormalization Group}\/} (Springer-Verlag, Berlin)

\bibitem{Metzner11}
Metzner W, Salmhofer M, Honerkamp C, Meden V and Sch\"onhammer K 
2012 {\em Rev. Mod. Phys.\/} in press 
(\eprint{arXiv:1105.5289})

\bibitem{Braun11}
Braun J 2011  (\textit{Preprint} \eprint{arXiv:1108.4449})

\bibitem{Schoenhammer01}
Sch\"onhammer K 2001 Unpublished

\bibitem{Meden03}
Meden V 2003 {Lecture notes on the Functional Renormalization Group}
  http://web.physik.rwth-aachen.de/ $\sim$meden/funRG/skriptumrg.pdf

\bibitem{Pawlowski10}
Pawlowski J~M 2010 {An introduction to the Functional RG \& applications to
  gauge theories} http://www.thphys.uni-heidelberg.de/$\sim$pawlowsk/
  NPgauge/FRG-Combo1.pdf

\bibitem{Salmhofer07}
Salmhofer M 2007 {\em Ann. Phys. (Leipzig)\/} {\bf 16} 171--206 

\bibitem{Schuetz05}
 Even if a model system is exactly solvable, it is still a nontrivial task to obtain this solution within the framework of the FRG. For an exact solution of the Tomonaga-Luttinger model, a well-known interacting many-body system in one dimensions, see Sch{\"u}tz F, Bartosch L and Kopietz P 2005 {\em Phys. Rev. B\/} {\bf 72} 035107

\bibitem{Delamotte04b}
For an elementary introduction to the perturbative renormalisation group, we refer to Delamotte B 2004 {\em Am. J. Phys.\/} {\bf 72} 170

\bibitem{Rossi98}
Rossi P, Campostrini M and Vicari E 1998 {\em Phys. Rep.\/} {\bf 302} 143--209

\bibitem{Arfken05}
Arfken G~B and Weber H~J 2005 {\em Mathematical methods for physicists\/} 6th
  ed (Academic Press, New York)

\bibitem{ZinnJustin02}
Zinn-Justin J 2002 {\em {Quantum Field Theory and Critical Phenomena}\/} 4th ed
  (Clarendon Press, Oxford)

\bibitem{Auerbach94}
Auerbach A 1994 {\em {Interacting Electrons and Quantum Magnetism}\/}
  (Springer-Verlag, New York)

\bibitem{Hedden04}
For a study of the {\em quantum} anharmonic oscillator, see Ref.~\cite{Gies06} or Hedden R, Meden V, Pruschke T and Sch\"{o}nhammer K 2004 {\em J. Phys.:
  Condens. Matter\/} {\bf 16} 5279--5296

\end{thebibliography}

\providecommand{\newblock}{}

\end{document}